\documentclass[a4paper,superscriptaddress,aps,12pt]{article}

\usepackage{graphicx,rotating,hyperref,slashed,amsmath,xcolor,amssymb,amsfonts,colortbl,cite,subcaption}
\usepackage[totalheight = 23cm, totalwidth = 17cm]{geometry}
\usepackage{simplewick,cancel,ulem}
\usepackage[export]{adjustbox}
\usepackage{float}

\definecolor{bluc}{cmyk}{1,1,0,0.1}
\definecolor{rossoCP3}{cmyk}{0,.88,.77,.40}
\definecolor{rosso}{cmyk}{0,1,1,0.4}
\definecolor{rossos}{cmyk}{0,1,1,0.55}
\definecolor{rossoc}{cmyk}{0,1,1,0.2}
\definecolor{verdes}{cmyk}{0.92,0,0.59,0.4}

\hypersetup{colorlinks, bookmarksopen, bookmarksnumbered,
citecolor=verdes, linkcolor=bluc, pdfstartview=FitH, urlcolor=rossos}

\newcommand{\calL}{{\cal L}}
\newcommand{\calO}{{\cal O}}
\newcommand{\calP}{{\cal P}}
\newcommand{\calR}{{\cal R}}

\newcommand{\calW}{{\cal W}}
\newcommand{\mpl}{m_{\rm Pl}}

\begin{document}

\begin{titlepage}

\rightline{\footnotesize{APCTP-Pre2025-021}}

\begin{center}

\vskip 3em

{\LARGE \bf 
Weakly model-independent determination
\vspace{0.2em}\\
of total expansion during inflation
}

\vskip 3em

{\large
Dayeong Choi$^{a,}$\footnote{\label{ref:footnote1}Equal contribution},
Subin Jeon$^{a,}$\textsuperscript{\ref{ref:footnote1}}
and 
Jinn-Ouk Gong$^{a,b,}$\footnote{Corresponding author}
}

\vskip 0.5cm

{\it
$^{a}$Department of Science Education,  Ewha Womans University, Seoul 03760, Korea
\\
$^{b}$Asia Pacific Center for Theoretical Physics, Pohang 37673, Korea
}

\end{center}

\vskip 1.2cm

\begin{abstract}

We study systematically the total expansion experienced by a certain perturbation mode during single-field inflation, not resorting to explicit models of inflation or reheating. By assuming that during the reheating stage the equation of state $w_\text{rh}$ can be written as a function of $e$-folds, the unknown dynamics during reheating parametrized by $w_\text{rh}$ is confined within a time integral so that any dependence on the models of inflation and reheating is isolated from model-independent contributions. Especially, the dependence on the reheating dynamics via $w_\text{rh}$ and the reheating temperature $T_\text{rh}$ is dominating. We give two illustrative examples of $w_\text{rh}$ to discuss its impacts on the total expansion, which can be different as much as 10 even for the same reheating temperature, depending on the shape of $w_\text{rh}$. We also discuss the profile degeneracy of $w_\text{rh}$, and argue when the degeneracy is lifted.

\end{abstract}

\end{titlepage}

\newpage

\section{Introduction}
\label{sec:intro}

The phase of accelerated expansion of the universe at very early times, cosmic inflation, provides homogeneity and isotropy of the observed universe as can be seen from the cosmic microwave background (CMB)~\cite{Guth:1980zm,Linde:1981mu,Albrecht:1982wi}. At the same time, during inflation tiny quantum vacuum fluctuations are generated and become the primordial perturbations. After inflation, these small inhomogeneities are responsible for structure formation by gravitational instability~\cite{Mukhanov:1981xt,Guth:1982ec,Hawking:1982cz,Starobinsky:1982ee,Bardeen:1983qw}. The properties of these primordial perturbations have been constrained over decades from the observations on the temperature anisotropies of the CMB and the large-scale distribution of galaxies, and are consistent with the inflationary predictions~\cite{Planck:2018vyg}.

To drive inflation, viz. to make the expansion of the universe accelerating, we need a very unusual matter content that has a strong negative pressure. One such a matter content is the cosmological constant, which however does not lead to a graceful exit from inflation. Instead, we usually introduce a scalar field whose potential energy is dominant over the kinetic one so that it behaves effectively as a cosmological constant, leading to an accelerated, quasi-exponential expansion of the universe. This scalar field that dominates the energy density of the universe during inflation is called the inflaton~\cite{Mukhanov:2005sc,Weinberg:2008zzc,Baumann:2022mni}.

%inflation is good

Meanwhile, the standard paradigm for physical cosmology, the hot big bang model, describes successfully the evolution of the universe starting from big bang nucleosynthesis (see e.g.~\cite{Steigman:2007xt,Cyburt:2015mya}) once the initial conditions consistent with current observations -- homogeneous and isotropic universe with $\calO(10^{-5})$ fluctuations -- are provided by inflation. This raises a very serious question: How does the universe become so hot after ``cold'' inflation? During inflation, the energy density of the universe is dominated by the inflaton. When the standard hot big bang evolution commences after inflation, relativistic matter contents such as photons occupy most of the energy density of the universe, along with tiny fraction of pressureless components like baryons and dark matter. This means the energy of the inflaton sector must be transferred to the usual matter contents with non-negative pressure, filling the universe with energetic particles. This process of energy transfer is called reheating (for reviews, see e.g.~\cite{Bassett:2005xm,Allahverdi:2010xz,Amin:2014eta}).

Thus, without reheating, after inflation the universe never becomes hot and thermalized\footnote{
Precisely speaking, this statement is true only if radiation is produced after inflation. If radiation exists during inflation in e.g. warm inflation~\cite{Berera:1995ie} the post-inflationary evolution including reheating stage could be different from the standard, ``cold'' inflation scenario (see e.g.~\cite{Berera:2023liv}).
}, not leading to the universe as observed. Reheating is an indispensable epoch that connects inflation and hot big bang. But as the universe, after reheating, is in thermal equilibrium, any model-dependent features that would have characterized the details of the reheating mechanism are swept away. This makes it very difficult to study the epoch of reheating. All we know is that the inflaton should decay and/or annihilate, and the final products are the standard model particles and dark matter. A standard, old approach is to introduce a decay width of the inflaton so that reheating proceeds perturbatively~\cite{Dolgov:1982th,Abbott:1982hn,Albrecht:1982mp}. However, since reheating epoch is so elusive, any non-perturbative process may well take place, such as preheating~\cite{Kofman:1994rk,Kofman:1997yn}, rescattering~\cite{Khlebnikov:1996mc,Khlebnikov:1996zt} and turbulence~\cite{Micha:2002ey,Micha:2004bv,Podolsky:2005bw}.

Connecting the reheating mechanism directly to the preceding inflationary epoch demands all the detailed properties of the inflaton -- how it is coupled to other species, what is the form of the potential after inflation around the minimum, and so on. However, as the identity of the inflaton itself is unknown, its properties are completely behind the veil. Thus an alternative approach is taken. Considering the simple inflation model with a quadratic potential, after inflation around the minimum of the potential the rapid oscillation of the inflaton makes it behave effectively as a pressuress matter, characterized by the equation of state $w \equiv p/\rho = 0$. Thus, phenomenologically we can adopt a specific equation of state during the reheating stage that encrypts a certain reheating mechanism~\cite{Turner:1983he}.

%reheating is unknown

In this article, we study systematically the contributions to the total expansion during inflation of a certain perturbation mode experiences. These contributions can be categorized into those dependent on and independent from the models of inflation and reheating dynamics. Only assuming that the equation of state during reheating can be written as a function of $e$-folds, we show that all the model dependence can be isolated from the model-independent contributions. This article is organized as follows. In Section~\ref{sec:expansion}, we consider the total expansion of a mode and separate individual contributions. By assuming that the equation of state during reheating is a function of $e$-folds, model-dependent contributions are sequestered from the model-independent ones. In Section~\ref{sec:model} we study the two model-dependent inputs, the endpoint of inflation and the equation of state during reheating. We present a more accurate estimate on the endpoint given an inflation model, along with a simple, more phenomenological parametrization. We give explicit examples for the equation of state during reheating, and show the total expansion of a mode as a function of model parameters explicitly. In Section~\ref{sec:degeneracy}, we argue that there exist degeneracies in different profiles of $w_\text{rh}$, the equation of state during reheating, as long as the average value of $w_\text{rh}$ is identical, and illustrate when and how the degeneracy is lifted and what information we can gain regarding reheating. Finally we conclude in Section~\ref{sec:conc}.

%a more convenient parametrization is given

\section{Total expansion a mode experiences}
\label{sec:expansion}

\subsection{Contributions to total expansion}

We begin with the ratio of comoving Hubble scale $a_kH_k = k$, when a certain perturbation mode with comoving wavenumber $k$ exited the horizon during inflation, to the present one $a_0H_0$. Here, we follow the minimal assumption that right after inflation the reheating stage follows. After reheating is complete the evolution of the universe can be described by the standard cosmological model including radiation and matter dominated epochs. Then, we may write the ratio $k/(a_0H_0)$ as~\cite{Liddle:2003as}:
\begin{equation}
\frac{k}{a_0H_0} 
=
\frac{a_kH_k}{a_0H_0}
=
\frac{a_k}{a_e} \frac{a_e}{a_\text{rh}} \frac{a_\text{rh}}{a_0} \frac{H_k}{H_0}
\, .
\end{equation}
Taking logarithm of both sides, and shuffling some terms gives
\begin{equation}
\label{eq:scale}
N_k
=
- \log \bigg( \frac{k}{a_0H_0} \bigg) - N_\text{rh} + \log \bigg( \frac{a_\text{rh}}{a_0} \bigg) + \log \bigg( \frac{H_k}{H_0} \bigg)
\, ,
\end{equation}
where $N_k = \log(a_e/a_k)$ and $N_\text{rh} = \log(a_\text{rh}/a_e)$ are the numbers of $e$-folds respectively from the moment during inflation the mode with comoving momentum $k$ crosses the horizon till the end of inflation, and from the end of inflation till radiation dominated epoch begins, viz. the end of reheating stage.

\begin{enumerate}

\item $N_k$: To explain the observed homogeneity and isotropy of CMB, we demand that this number be larger than a certain value in any successful inflation model. Assuming that inflation is driven by a single inflaton field $\phi$, $N_k$ can be estimated by trading time with the value of $\phi$ as
\begin{equation}
N_k = \int_k^e H dt
= \int_{\phi_k}^{\phi_e} \frac{H}{\dot\phi} d\phi
\, ,
\end{equation}
where we have assumed that $\phi = \phi(t)$. Thus, to determine $N_k$, we need two inputs. On one hand, we need to specify exactly the two boundary values of $\phi$: $\phi_k$ when the $k$-mode crosses the horizon, and $\phi_e$ when inflation ends. On the other hand, we should also be able to write both $H$ and $\dot\phi$ as functions of $\phi$. However, it is usually impossible to find the full analytic solution for $\phi(t)$ as we need to solve the highly non-trivial equation of motion for $\phi$. Instead, we usually resort to the slow-roll approximation to gain analytic control. Furthermore, solving for $\phi(t)$ to determine two boundaries and to find $H$ and $\dot\phi$ means that we specify the model of inflation.

\item $\log\big[k/(a_0H_0)\big]$: Since $a_0$ is arbitrarily normalized, conventionally we set $a_0 = 1$. Then, the reference wavenumber $k$ corresponds to the present value. Regarding the CMB observations, $k = \calO(10^{-3} - 10^{-2})$ Mpc$^{-1}$. The present Hubble parameter $H_0$ can be determined by various observations on, for example, CMB~\cite{Planck:2018vyg}. Thus this term can be pinned down without any ambiguity up to observational errors in $H_0$.

\item $N_\text{rh}$: This number measures the expansion of the universe starting from the end of inflation until reheating ends so that the universe becomes dominated by radiation. As $N_\text{rh}$ depends on the detail of the reheating mechanism that remains mostly elusive, we usually resort to an effective and/or phenomenological approach. One possible way is to make use of $w_\text{rh}$ and to assume that it is a constant, e.g. $w_\text{rh} = (n-2)/(n+2)$ for $V(\phi) \propto \phi^n$ when averaged over oscillation cycles~\cite{Martin:2006rs}. This leads to the following relation between the energy densities at the end of inflation $\rho_e$ and at the end of reheating stage $\rho_\text{rh}$:
\begin{equation}
N_\text{rh} = \frac{1}{3(1+w_\text{rh})} \log \bigg( \frac{\rho_e}{\rho_\text{rh}} \bigg) \, .
\end{equation}

\item $\log(a_\text{rh}/a_0)$: Since radiation domination, we can follow the standard thermal history of the universe. Assuming entropy conservation after reheating, $a_\text{rh}/a_0$ can be written as (see e.g.~\cite{Kolb:1990vq})
\begin{equation}
\label{eq:arh-a0}
\frac{a_\text{rh}}{a_0} = \bigg( \frac{11}{43} g_{*S} \bigg)^{-1/3} \frac{T_0}{T_\text{rh}} \, ,
\end{equation}
where $g_{*S}$ is the effective number of light species for entropy at the moment of reheating. Thus, in determining $\log(a_\text{rh}/a_0)$, the most unclear factor is the reheating temperature $T_\text{rh}$. This is closely related to $N_\text{rh}$, as once it is figured out we can calculate immediately, starting from $\rho_e$, the energy density at the end of reheating $\rho_\text{rh}$, which gives $T_\text{rh}$ by
\begin{equation}
\label{eq:Trh}
\rho_\text{rh} = \frac{\pi^2}{30} g_* T_\text{rh}^4 \, ,
\end{equation}
where $g_*$ is the effective number of relativistic species at the moment of reheating, and is not necessarily the same as $g_{*S}$. Both $g_*$ and $g_{*S}$ depend on temperature, and assuming the validity of the standard model of particle physics up to $\calO(100)$ GeV, $g_* = 106.75$.

\item $\log(H_k/H_0)$: We need to know the value of the Hubble parameter when the $k$-mode exits the horizon. This can be fixed in two different ways. First, from $N_k$, we know $\phi_k$ and thus can determine $H_k$ for a given model of inflation. Or, from the value of the scalar power spectrum $\calP_\calR$ which is strongly constrained on the CMB scales, we can write $H_k$ in terms of $\calP_\calR$ [see \eqref{eq:Hk-PR}].

\end{enumerate}

\subsection{Sequestering model dependence}

As we have seen, estimating $N_k$ exactly requires that we specify the model of inflation and solve the equations numerically. Even if we are to make use of the right-hand side of \eqref{eq:scale}, we still need to know the model of inflation via $\rho_e$ and $H_k$, and additionally, on top of inflation, the detail of the reheating process through $N_\text{rh}$ and $a_\text{rh}$. Thus one may be tempted to conclude that at least without specifying the inflation model it is difficult to estimate $N_k$. But we can improve the  situation as follows. If the universe is dominated by a canonical scalar field, viz. the inflaton, the acceleration of the scale factor is determined by the following equation, with $\rho_\phi = \dot\phi^2/2+V$ and $p_\phi = \dot\phi^2/2-V$:
\begin{equation}
\frac{\ddot{a}}{a} = -\frac{\rho_\phi + 3p_\phi}{6\mpl^2} = \frac{\dot\phi^2 - V}{3\mpl^2} \, .
\end{equation}
During slow-roll phase, $\dot\phi^2 \ll V$ so that $\ddot{a} > 0$, i.e. the expansion is accelerating and the universe undergoes inflation. Meanwhile, inflation ends at $\phi=\phi_e$ when the following condition is satisfied so that $\ddot{a} = 0$:
\begin{equation}
\label{eq:end-condition}
\dot\phi_e^2 = V(\phi_e) \, .
\end{equation}
Thus our first observation is that, from \eqref{eq:end-condition}, at the end of inflation $w_e = -1/3$:
\begin{equation}
\label{eq:eos-end}
w_e = \frac{p_\phi}{\rho_\phi} \bigg|_e = \frac{\dot\phi^2/2-V}{\dot\phi^2/2+V} \bigg|_e
= -\frac{1}{3}
\, .
\end{equation}

At the beginning of radiation domination, viz. at the end of reheating epoch, the equation of state is that of radiation: $w_r = 1/3$. That is, at both ends of the reheating stage, the values of the equation of state are fixed, $-1/3$ at the beginning and $1/3$ at the end. Thus we expect that any realistic equation of state during reheating begins from $-1/3$, changes smoothly, and ends up at the value of $1/3$.\footnote{
Here, we assume that during the reheating state $w_\text{rh}$ does not become smaller than $-1/3$. If so, until $w_\text{rh}>-1/3$ again the universe undergoes yet another inflationary stage. 
} 
Now we assume that $w_\text{rh}$ can be written as a function of $e$-folds, with $N=0$ at the beginning and $N=N_\text{rh}$ at the end of the reheating stage. Then $w_\text{rh}(N=0) = -1/3$ and $w_\text{rh}(N=N_\text{rh}) = 1/3$. From the continuity equation, we can solve for the energy density at the end of reheating, $\rho_\text{rh}$, with a general equation of state during reheating which is a function of the normalized $e$-folds $n\equiv N/N_\text{rh}$:
\begin{equation}
\label{eq:rho-rh}
\rho_\text{rh} = \rho_e \exp \left\{ -3N_\text{rh} \int_0^1 \big[ 1+w_\text{rh}(n) \big] dn \right\} \, .
\end{equation}
Thus, we can isolate the effects of the model-dependent equation of state during reheating within the time integration. On general ground, the integration of $w_\text{rh}(n)$ can be performed analytically and/or numerically:
\begin{equation}
\label{eq:calW}
\int_0^1 [1+w_\text{rh}(n)] dn
=
1 + \int_0^1 w_\text{rh}(n) dn
\equiv
1+ \calW(\alpha)
\, ,
\end{equation}
where $\alpha$ collectively denotes a set of parameters on which the evolution of $w_\text{rh}(n)$ depends. We expect that $\calW(\alpha)$ is likely to be a number of $\calO(1)$. We will consider explicit examples of $w_\text{rh}(n)$ later.

Now, using \eqref{eq:arh-a0}, we can write \eqref{eq:scale} as
\begin{equation}
\label{eq:scale2}
N_k
=
- \log \bigg( \frac{k}{a_0H_0} \bigg) - N_\text{rh} - \frac{1}{3} \log \bigg( \frac{11}{43}g_{*S} \bigg) 
+ \log \bigg( \frac{T_0}{H_0} \bigg) + \log \bigg( \frac{H_k}{T_\text{rh}} \bigg)
\, .
\end{equation}
Further, using \eqref{eq:Trh}, \eqref{eq:rho-rh} and \eqref{eq:calW}, the number of $e$-folds during reheating can be written as
\begin{align}
\label{eq:Nrh}
N_\text{rh}
& =
\frac{1}{3(1+\calW)} \log \bigg( \frac{\rho_e}{\rho_\text{rh}} \bigg)
\nonumber\\
& =
\frac{1}{3(1+\calW)} \bigg[ \log \bigg( \frac{\rho_e}{\mpl^4} \bigg)
- \log \bigg( \frac{\pi^2}{30}g_* \bigg)
- 4 \log \bigg( \frac{T_\text{rh}}{\mpl} \bigg) \bigg]
\, .
\end{align}
Note that $N_\text{rh}\geq0$ gives the upper bound on $T_\text{rh}/\mpl$ by demanding that the terms in the square brackets of \eqref{eq:Nrh} be non-negative.\footnote{
As long as we assume $w_\text{rh}>-1/3$, the prefactor $1+\calW$ is always positive. For the extreme case of $w_\text{rh}=-1/3$ until the end of reheating, $\calW = -1/3$.
} $T_\text{rh}$ is saturated when $N_\text{rh} = 0$, viz. instantaneous reheating. Meanwhile, from the slow-roll approximation which is likely to be very effective for the scales corresponding to the CMB observations, we can write the power spectrum of the scalar perturbation as
\begin{equation}
\calP_\calR = \bigg( \frac{H}{2\pi} \bigg)^2 \bigg( \frac{H}{\dot\phi} \bigg)^2
= \frac{2}{\pi^2r} \frac{H^2}{\mpl^2}
\, ,
\end{equation}
where $r = 16\epsilon$, with $\epsilon \equiv -\dot{H}/H^2$ being the slow-roll parameter, is the tensor-to-scalar ratio for single-field inflation. This enables us to replace $H_k$ with the observationally constrained values of $\calP_\calR$ and $r$ as~\cite{Dai:2014jja}
\begin{equation}
\label{eq:Hk-PR}
\frac{H_k}{\mpl} = \sqrt{\frac{r\calP_\calR}{2}} \pi
\, .
\end{equation}
Here, we may include higher-order slow-roll corrections to $\calP_\calR$~\cite{Stewart:1993bc,Gong:2001he} as well as $r$~\cite{Martin:2006rs,Creminelli:2014oaa}. But such corrections do not lead to significant changes, as $N_k$ is dependent on $\calP_\calR$ and $r$ only logarithmically. So it is sufficient to take the leading slow-roll results for $\calP_\calR$ and $r$.

Combining \eqref{eq:scale2}, \eqref{eq:Nrh} and \eqref{eq:Hk-PR}, we find $N_k$ as
\begin{align}
\label{eq:Nk}
N_k
& =
- \log \bigg( \frac{k}{a_0H_0} \bigg) 
- \frac{1}{3} \log \bigg( \frac{11}{43} g_{*S} \bigg) 
+ \log \left( \frac{T_0}{H_0} \right) + \log \left( \sqrt{\frac{r\calP_\calR}{2}} \pi \right) 
\nonumber\\
& \quad
+ \frac{1}{3(1+\calW)} \log \bigg( \frac{\pi^2}{30}g_* \bigg)
- \frac{4}{3(1+\calW)} \log \Bigg( \frac{\rho_e^{1/4}}{\mpl} \Bigg)
+ \frac{1-3\calW}{3(1+\calW)} \log \bigg( \frac{T_\text{rh}}{\mpl} \bigg) 
\, .
\end{align}
Notice that the terms in the first line of \eqref{eq:Nk} can be all completely fixed within our observational errors and theoretical uncertainties, and are nearly independent of the detailed dynamics during inflation and reheating. Meanwhile, the terms in the second line of \eqref{eq:Nk} -- especially the last two terms -- contain all the model dependence of inflation and reheating via $\calW(\alpha)$, $\rho_e$ and $T_\text{rh}$. Note that if $\calW=1/3$, the last term vanishes and $T_\text{rh}$ does not affect $N_k$. As can be read from \eqref{eq:calW}, this is the case if simply $w_\text{rh} = 1/3$, i.e. during reheating the universe is dominated by radiation. But even if $w_\text{rh}\neq1/3$, we may say the universe is effectively radiation-dominated in the sense that $\calW=1/3$ so that $N_k$ is independent of $T_\text{rh}$.

\section{Model-dependent inputs}
\label{sec:model}

\subsection{Endpoint of inflation}

The energy density at the end of inflation, $\rho_e$, is certainly model-dependent. Usually, it is identified as the value of the potential energy when the potential slow-roll parameter $\epsilon_V \equiv \mpl^2 (V'/V)^2/2$ with $V' = dV/d\phi$ becomes 1. For a given model of inflation, however, we can determine the endpoint of inflation without resorting to the slow-roll approximation as follows. Assuming no interaction with any other matter contents during inflation, $\phi$ satisfies the following equation of motion:
\begin{equation}
\label{eq:phi-eom}
\ddot\phi + 3H\dot\phi + V' = 0 \, .
\end{equation}
Here, we can define the second slow-roll parameter $\eta$ in such a way that
\begin{equation}
\eta \equiv \frac{\dot\epsilon}{H\epsilon} 
= -\frac{\ddot{H}}{H^3\epsilon} + 2\epsilon
\, ,
\end{equation}
so we can replace $\ddot\phi$ with $\dot\phi$ and slow-roll parameters:
\begin{equation}
\ddot\phi 
= 
-\frac{\mpl^2}{\dot\phi} \ddot{H}
=
-\frac{\mpl^2}{\dot\phi} H^3 \epsilon (2\epsilon - \eta)
\, ,
\end{equation}
and \eqref{eq:phi-eom} can be written as
\begin{equation}
\label{eq:phi-eom2}
-\frac{\mpl^2}{\dot\phi} H^3 \epsilon ( 2\epsilon - \eta ) + 3H\dot\phi + V_\phi = 0 
\, .
\end{equation}

Now, at $\phi_e$ where \eqref{eq:end-condition} is satisfied, we can write $\dot\phi_e$ in terms of $V(\phi_e) \equiv V_e$ as
\begin{equation}
\dot\phi_e = -\sqrt{V_e}
\, ,
\end{equation}
where we have assumed that during inflation $\phi > 0$ and it approaches to 0, hence $\dot\phi < 0$. Then from \eqref{eq:phi-eom2} we can find that at $\phi_e$ the following relation is exactly satisfied:
\begin{equation}
\frac{V_e}{V_e'} = \frac{2\sqrt{2}}{4+\eta_e} \mpl
\, ,
\end{equation}
where $\eta_e$ is the value of $\eta$ at $\phi_e$ and is typically a constant of $\calO(1)$. To have a more concrete idea, let us consider a general power-law potential, $V(\phi) = m^{4-n}\phi^n$. This gives directly %the value of $\phi_e$ as:
\begin{equation}
\label{eq:phie-estimate}
\frac{\phi_e}{\mpl} = \frac{2\sqrt{2}n}{4+\eta_e} 
\, .
\end{equation}
The usual endpoint based on the potential slow-roll parameter $\epsilon_V \big|_e = 1$ is
\begin{equation}
\label{eq:SRphie}
\frac{\phi_e}{\mpl} =  \frac{n}{\sqrt{2}} \, .
\end{equation}
In Table~\ref{table:endpoint} for the power-law potential we compare the numerical value of $\phi_e/\mpl$ with \eqref{eq:phie-estimate} for different values of $\eta_e$, along with the usual slow-roll approximation result \eqref{eq:SRphie}. Even if including interactions with other species, e.g. perturbative decay of $\phi$ as
\begin{equation}
\label{eq:eom-decay}
\ddot\phi + 3H\dot\phi + V_\phi = -\Gamma\dot\phi \, ,
\end{equation}
with $\Gamma$ being a constant decay width, the value of $\phi_e$ is not changed appreciably as can be seen in Table~\ref{table:endpoint}.

\begin{table}[t]
\begin{center}
\begin{tabular}{c||c|c|c}
$n$ & 2 & 3 & 4
\\\hline\hline
Numerical values & 1.00938 & 1.66602 & 2.33939
\\\hline
\eqref{eq:phie-estimate} ($\eta_e=0.5$) & 1.25708 & 1.88562 & 2.51416
\\\hline
\eqref{eq:phie-estimate} ($\eta_e=1.0$) & 1.13137 & 1.69706 & 2.26274
\\\hline
\eqref{eq:phie-estimate} ($\eta_e=1.5$) & 1.02852 & 1.54278 & 2.05704
\\\hline
\eqref{eq:SRphie} & 1.41421 & 2.12132 & 2.82843 
\\\hline\hline
$\Gamma = 0.1m$ & 0.97188 & 1.66593 & 2.33939
\\\hline
$\Gamma = 0.01m$ & 1.00556 & 1.66601 & 2.33939
\\\hline
$\Gamma = 0.001m$ & 1.00899 & 1.66602 & 2.33939
\end{tabular}
\caption{Comparison of $\phi_e/\mpl$ between numerical results, \eqref{eq:phie-estimate} with different values of $\eta_e$, and the standard slow-roll approximation \eqref{eq:SRphie} based on the power-law potential $V(\phi) = m^{4-n}\phi^n$. Typically $\eta_e = 1$ gives reasonably good approximation for $\phi_e$. We also present $\phi_e/\mpl$ obtained numerically with different values of the decay width $\Gamma$ as given by \eqref{eq:eom-decay}.}
\label{table:endpoint}
\end{center}
\end{table}

Although we have discussed a refined estimate for $\phi_e$ for a given model of inflation, more phenomenologically we may simplify a lot as follows. Denoting the ratio of the energy density during inflation corresponding to the CMB scales to $\rho_e$ as a constant $\beta$, i.e. $\beta \equiv \rho_k/\rho_e$, we can write $\rho_e$ in terms of $H_k$ and thus in turn $\calP_\calR$ and $r$ as 
\begin{equation}
\label{eq:rhoe-ph}
\log \left( \frac{\rho_e^{1/4}}{\mpl} \right) 
= \log \left( \beta^{-1/4} \frac{\rho_k^{1/4}}{\mpl} \right)
= -\frac{1}{4} \log \bigg(\frac{\beta}{3}\bigg) + \frac{1}{2} \log \left( \sqrt{\frac{r\calP_\calR}{2}} \pi \right)
\, .
\end{equation}
Note that we do not define $\beta$ as the ratio of $\rho_\text{rh}$ to $\rho_e$ as in~\cite{Gong:2015qha}, because then $\beta$, $N_\text{rh}$ and $T_\text{rh}$ are interchangeable. The constant $\beta$ as defined here characterizes a large class of inflation models. For example, for a large-field power-law potential model typically we have $\beta \gtrsim \calO(100)$, while for hybrid inflation or hilltop inflation $\beta = \calO(1)$. However, as the dependence on $\beta$ is logarithmic, the contributions to $N_k$ of different models of inflation are not very different: The first term of \eqref{eq:rhoe-ph}, which in this way solely contains the dependence on the inflation model, gives $-0.300993$, $-0.876639$ and $-1.45229$ for $\beta=10$, 100 and 1000 respectively.

\subsection{Equation of state during reheating}

\begin{figure}[t]
\begin{center}
 \makebox[0pt]{\includegraphics[width=1.2\textwidth]{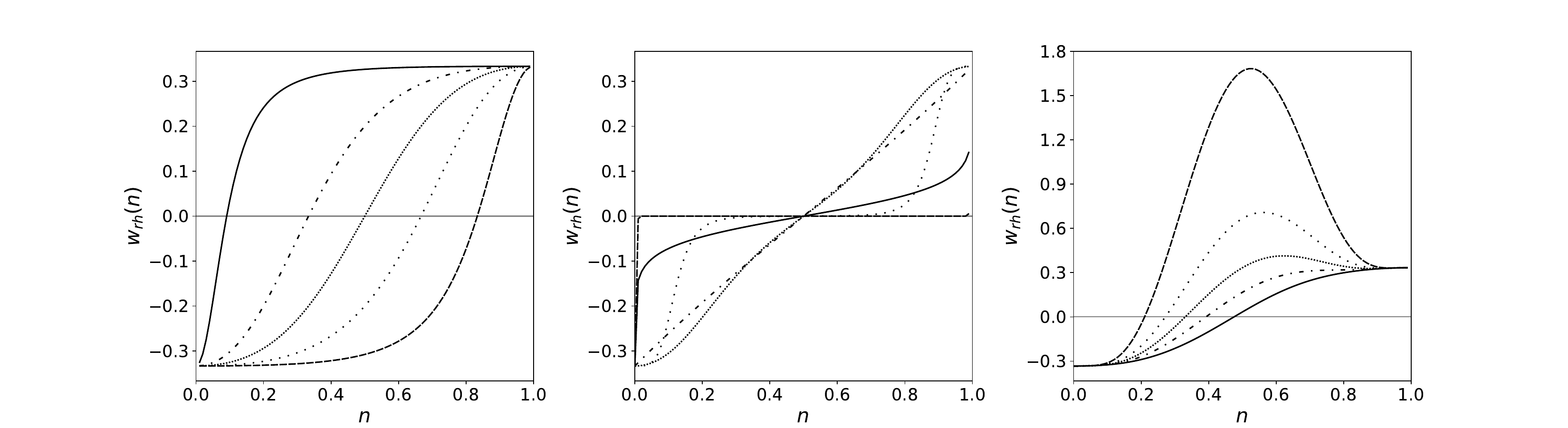}}
\caption{Plots of $w_\text{rh}$ as a function of $n\equiv N/N_\text{rh}$ given respectively by (left) \eqref{eq:w-rh1}, (middle) \eqref{eq:w-rh2} and (right) \eqref{eq:w-rh3}. Here, (solid) $\alpha=0.1$, (dot-dashed) 0.5, (dense dot) 1, (sparse dot) 2 and (dashed) 5 respectively.}
\label{fig:w-rh}
\end{center}
\end{figure}

As mentioned, we do not have any standard model of reheating process. So usually we have to make a set of reasonable assumptions and take phenomenological approaches to reheating. Here, we only assume that $w_\text{rh}$ can be written as a function of the normalized $e$-folds during reheating $n \equiv N/N_\text{rh}$, which encodes the detailed process during reheating as a smooth function starting from $-1/3$ and ending at $1/3$. As one example for such an equation of state during reheating, we model $w_\text{rh}(n)$ as the following:
\begin{equation}
\label{eq:w-rh1}
w_\text{rh}(n) = \frac{1-\alpha^2 \big( n^{-1}-1 \big)^2}{3 \Big[ 1+\alpha^2 \big( n^{-1}-1 \big)^2 \Big]} \, ,
\end{equation}
where $\alpha>0$ is a constant. In the left panel of Figure~\ref{fig:w-rh}, we present the evolution of $w_\text{rh}$ for several different values of $\alpha$. Further, we can perform the integral \eqref{eq:calW} analytically to find $\calW(\alpha)$ as
\begin{equation}
\label{eq:calW-model}
\calW(\alpha)
=
\frac{1}{3(1+\alpha^2)^2} \Big[ (1-\alpha^4) - \alpha\pi (1-\alpha^2) - 4\alpha^2\log\alpha \Big]
\, .
\end{equation}
For $\alpha=0$, the value of $\calW$ is 1/3, monotonically decreases and approaches $-1/3$ for $\alpha\to\infty$. Note that for $\alpha=1$, $w_\text{rh}(n)$ is an odd function around $n=1/2$ thus $\calW = 0$.

As another example, consider the following $w_\text{rh}(n)$:
\begin{equation}
\label{eq:w-rh2}
w_\text{rh}(n) 
=
\frac{1}{6} \bigg\{ \tanh \bigg[ \alpha \bigg( \log \bigg( \frac{n}{1-n} \bigg) - \alpha \bigg) \bigg]
+ \tanh \bigg[ \alpha \bigg( \log \bigg( \frac{n}{1-n} \bigg) + \alpha \bigg) \bigg] \bigg\}
\, ,
\end{equation}
where $\alpha > 0$. We show \eqref{eq:w-rh2} as a function of $n$ in the middle panel of Figure~\ref{fig:w-rh} for different values of $\alpha$. In this case, since $w_\text{rh}(n)$ is symmetric around $n=1/2$, simply $\calW = 0$ irrespective of the value of $\alpha$. Then during the reheating epoch the energy density scales as that of pressureless matter, although the behaviour of $w_\text{rh}$ could be substantially different from $w_m = 0$.

For these two examples, $w_\text{rh}(n)$ remains between $-1/3$ and $1/3$. But this needs not be the case if the universe after inflation experiences an exotic period. For example, during kinetic domination the equation of state becomes 1. As such an example with $w_\text{rh} \geq 1/3$, we consider the following:
\begin{equation}
\label{eq:w-rh3}
w_\text{rh}(n) = \frac{2n-1}{3\big(2n^2-2n+1\big)}
+ \frac{\alpha}{3} \exp \bigg[ -\log^2 \bigg( \frac{1-n}{n} \bigg) \bigg] 
\, .
\end{equation}
As shown in the right panel of Figure~\ref{fig:w-rh}, \eqref{eq:w-rh3} exhibits at $n=1/2$ a bump the height of which is set by $\alpha$ and can exceeds $1/3$. We found it is difficult to perform the integral of \eqref{eq:w-rh3} analytically, but numerically it is easily done and $\calW$ is linearly proportional to $\alpha$, with $\calW(\alpha=0) = 0$.

One may, by taking into account the oscillating inflaton field around the minimum of the potential, include sinusoidal modulations that diminish at both ends into $w_\text{rh}$ as the following:
\begin{equation}
\includegraphics[valign=c,width=0.18\textwidth]{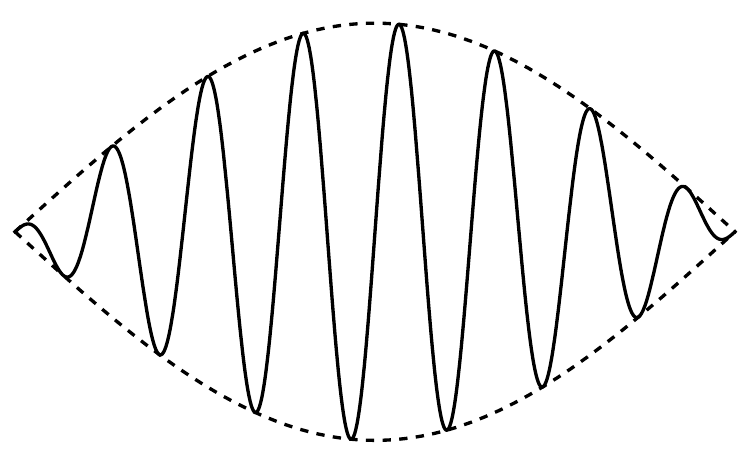} ~
+ ~ \centering{\includegraphics[valign=c,width=0.18\textwidth]{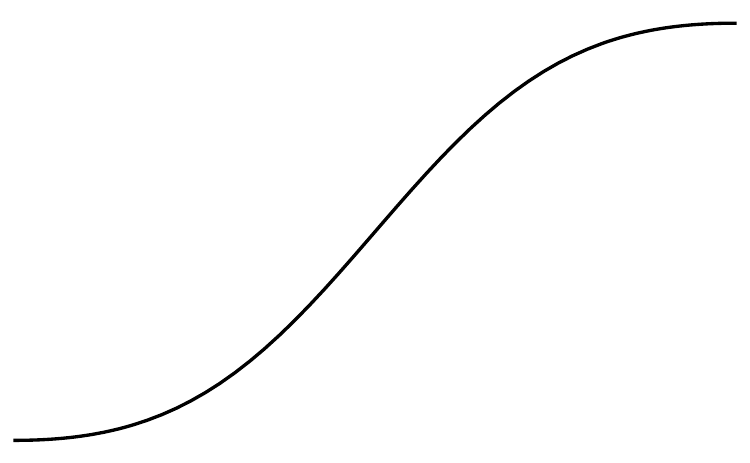}} ~
= ~ \centering{\includegraphics[valign=c,width=0.18\textwidth]{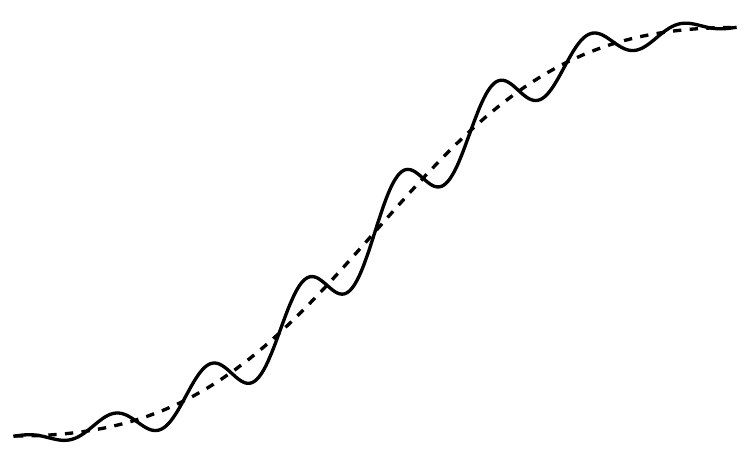}}
\end{equation}
Such a (slightly) oscillating behavior might be a more realistic approximation for $w_\text{rh}$. However, integrating such oscillations over the reheating period typically vanishes. This leads to little change in $\calW$ even if we include sinusoidal modulations. Thus, we do not consider any diminishing oscillations in $w_\text{rh}$, not because such oscillations are unrealistic but because they do not lead to any appreciable difference.

Thus, as examples for a non-trivial equation of state during reheating epoch, we take \eqref{eq:w-rh1} and \eqref{eq:w-rh3} and see the corresponding number of $e$-folds a certain $k$-mode has gained during inflation. For the model-independent contributions, we take the following values from observations and theoretical estimates based on the standard model of particle physics:
\begin{table}[H]
\begin{center}
\begin{tabular}{c|c|c|c|c}
$a_0$ & $H_0$ & $g_* = g_{*S}$ & $T_0$ & $\calP_\calR$
\\\hline\hline
1 & $67.4 \text{ km s}^{-1} \text{Mpc}^{-1}$ & 106.75 & 2.725 K & $2.0968\times10^{-9}$
\end{tabular}
\caption{Input values for the model-independent contributions in \eqref{eq:Nk}.}
\label{table:values}
\end{center}
\end{table}

\noindent
As a fiducial value of the tensor-to-scalar ratio, we take $r = 10^{-3}$. We consider two representative wavenumbers, $k = 0.05$ Mpc$^{-1}$ and $k = 0.002$ Mpc$^{-1}$. The model-independent contributions in \eqref{eq:Nk}, viz. the first four terms, give 48.1115 for $k = 0.05$ Mpc$^{-1}$, and 51.3304 for $k = 0.002$ Mpc$^{-1}$ respectively. Then, for a given model of inflation, i.e. for a fixed value of $\beta$, \eqref{eq:Nk} with \eqref{eq:rhoe-ph} and $w_\text{rh}(n)$ becomes a function of $\alpha$ and $T_\text{rh}$.

\begin{figure}[ht]
\begin{center}
 \makebox[0pt]{\includegraphics[width=1.1\textwidth]{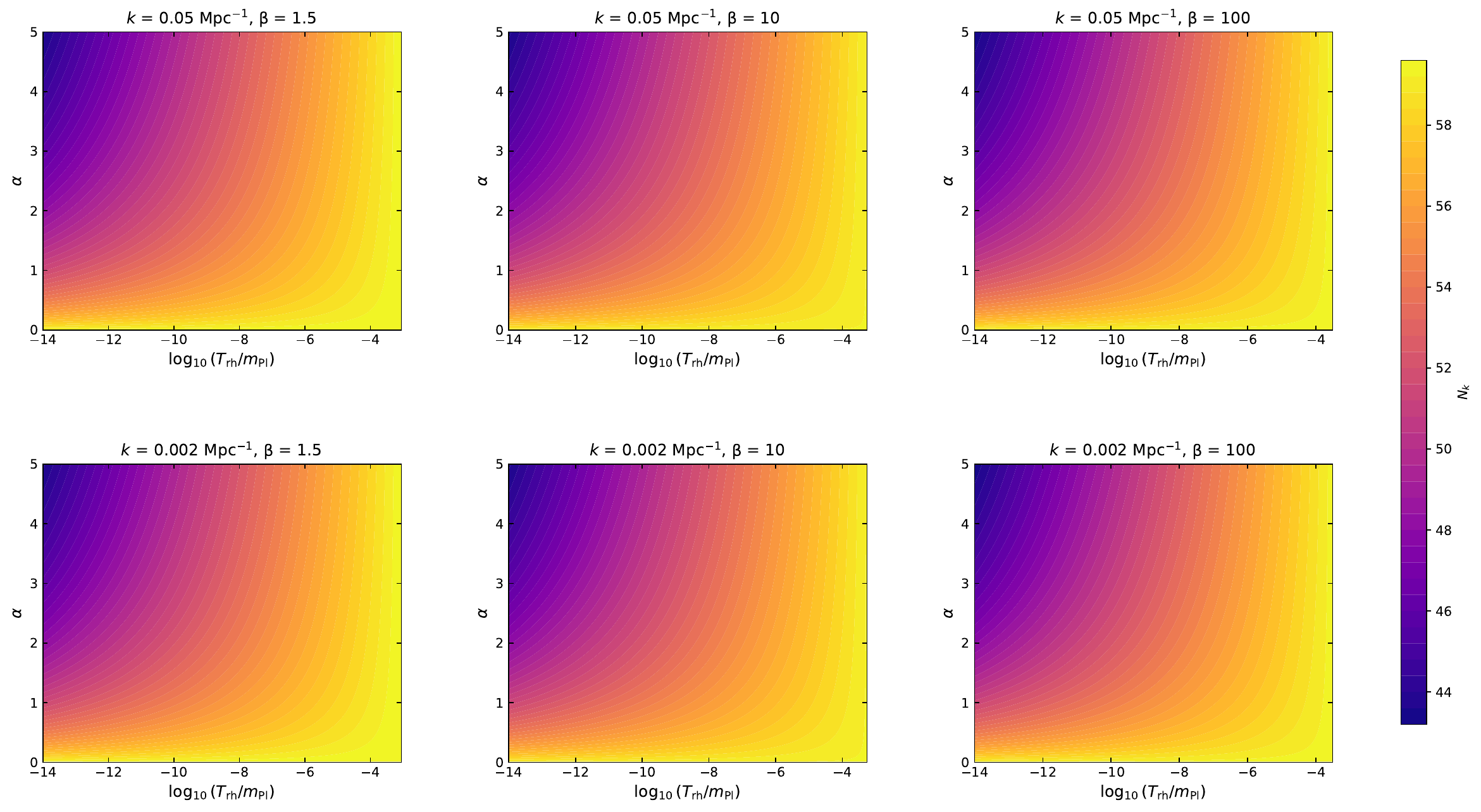}}
 \\\vspace{0.5em}
 \makebox[0pt]{\includegraphics[width=1.1\textwidth]{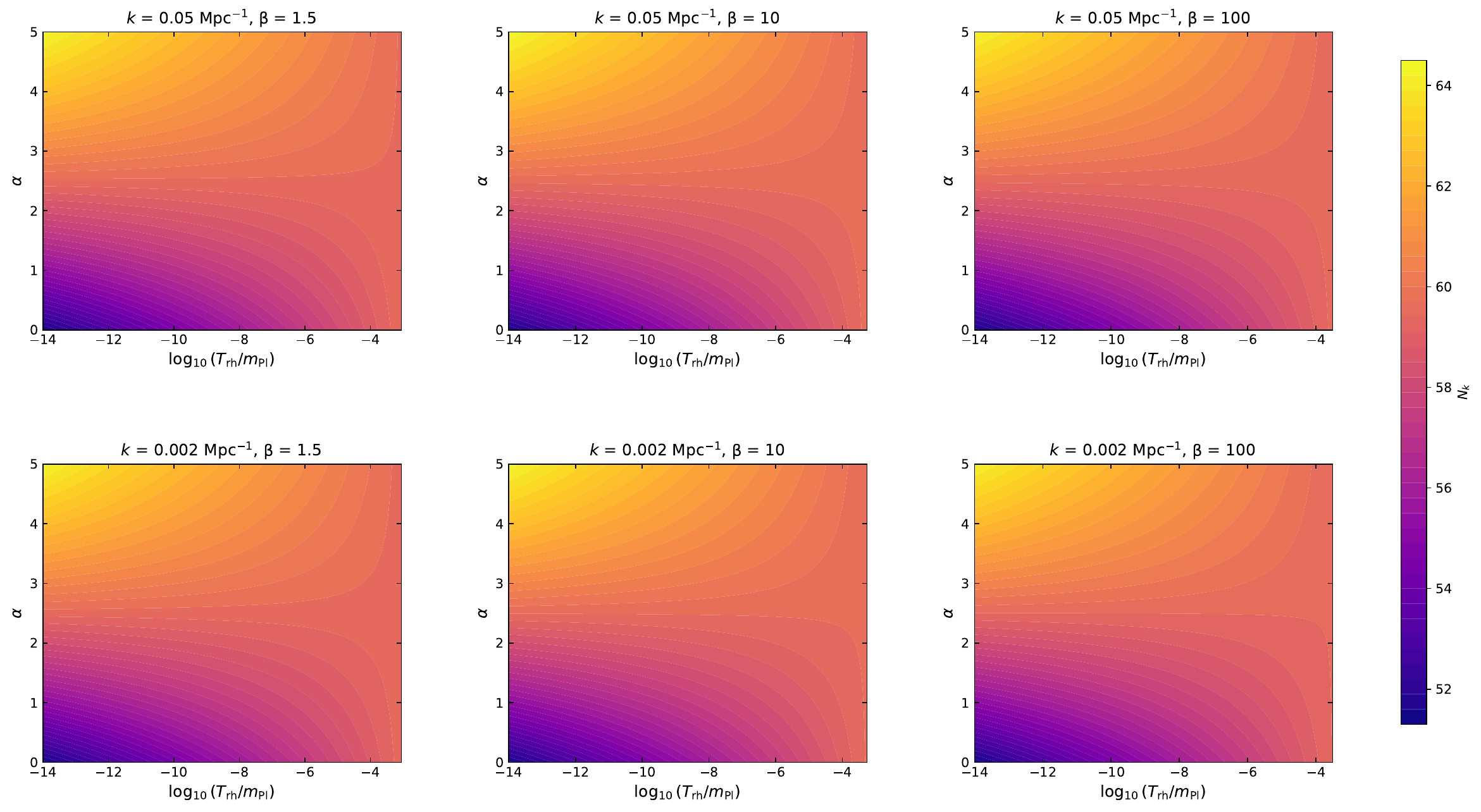}}
\caption{Plots of $N_k$ for (upper panels) \eqref{eq:w-rh1} and (lower panels) \eqref{eq:w-rh3} as a function of (horizontal axis) $T_\text{rh}/\mpl$ and (vertical axis) $\alpha$. In both upper and lower panels, the reference scale is set to be (upper rows) $k=0.05$ Mpc$^{-1}$ and (lower rows) $k=0.002$ Mpc$^{-1}$. The value of $\beta$ is set to be (left panels) $\beta = 1.5$, (middle panels) $\beta=10$ and (right panels) $\beta=100$. $T_\text{rh}/\mpl$ is presented up to the value that gives $N_\text{rh}=0$.}
\label{fig:Nk-model1} 
\end{center}
\end{figure}

In Figure~\ref{fig:Nk-model1}, we show \eqref{eq:Nk} as a function of $T_\text{rh}/\mpl$ and $\alpha$ for given values of $\beta$, with the equation of state during inflation being (upper panel) \eqref{eq:w-rh1} and (lower panel) \eqref{eq:w-rh3}. In the upper panels where we have considered \eqref{eq:w-rh1}, the larger $\alpha$ is, the smaller $N_k$ is. This is because, as can be read from the left panel of Figure~\ref{fig:w-rh}, a larger value of $\alpha$ gives a slower progress of reheating, giving rise to a lower value of $N_k$. Note that as $\alpha\to0$, it is no different from instantaneous reheating so there is no constraint on $N_k$ from the possible value of $T_\text{rh}/\mpl$. But generally, depending on the value of $\alpha$, even for the same reheating temperature $N_k$ could be different as large as $\calO(10)$.  In the lower panels where the equation of state during reheating is taken as \eqref{eq:w-rh3}. As the equation of state exceeds $1/3$ for a larger value of $\alpha$, we then have a larger values of $N_k$. This happens when $\calW = 1/3$ so that effectively $\calW$ is identical to the case of instantaneous reheating. This occurs at $\alpha \approx 2.51171$ where, as can be seen, $N_k$ is independent of $T_\text{rh}/\mpl$. Across this line $\Delta{N}_k = \calO(10)$ for a wide range of $\alpha$.

\section{Degeneracy with the effective equation of state}
\label{sec:degeneracy}

%section: degeneracy with the effective equation of state
%- the formulation is blind to the profile of w(N)
%subsection: when the degeneracy is broken
%- when looking at other observables (eft language first, cases later)
%- when g_* varies [what's different from above (bg language, not pert), eft language]
%subsection: explicit example
%- mcmc with delta{N}_k propto alpha
%- step variation of g_*

Until now, we have considered the separation of the model-dependent and weakly independent contribution to the total expansion. As can be read from \eqref{eq:Nk}, the detail during the reheating stage is confined within $\calW$, with $\rho_e$ and $T_\text{rh}$ being the fixed boundaries at the beginning and end of the reheating stage respectively. Thus, if we define the {\it average} equation of state $w_\text{eff}$ as~(see e.g. \cite{Ellis:2015jpg})
\begin{equation}
w_\text{eff} \equiv %\frac{1}{t_\text{rh} - t_e} \int_{t_i}^{t_\text{rh}} w_\text{rh}(t) dt
\frac{1}{N_\text{rh}} \int_0^{N_\text{rh}} w_\text{rh}(N) dN
= \int_0^1 w_\text{rh}(n) dn
= \calW
\, ,
\end{equation}
there is no difference at all from just using $w_\text{eff}$ instead of $\calW$. That is, the information on the reheating stage contained in $N_k$ is only the average value of $w_\text{rh}$. More precisely speaking, given a model of inflation specified by $\rho_e$, assuming a standard thermal history with constant relativistic degrees of freedom $g_*$, the reheating contribution to $N_k$ depends only on the averaged equation of state $w_\text{eff}$ and the reheating temperature $T_\text{rh}$. In this case, different reheating profiles $w_\text{rh}(N)$ that share the same $w_\text{eff}$ are {\it exactly} degenerate in their prediction for $N_k$. Our parameter $\alpha$ that specifies the evolution of $w_\text{rh}$ during reheating therefore does not introduce an additional degree of freedom for itself in this case. However, this does not mean that using the full profile of $w_\text{rh}$ is useless. Rather, what we have seen above indicates that if we choose to compute $N_k$ alone, the information on the details of $w_\text{rh}$ are collapsed into the average value $w_\text{eff}$: $N_k$ alone is blind to the shape of reheating.

\subsection{When the degeneracy is broken}

Therefore, the shape of $w_\text{rh}$ becomes relevant when our assumptions are violated: During reheating the universe is described by a single barotropic fluid with a constant $g_*$, and we are only interested in $N_k$. Thus, there are largely two different ways to break the profile degeneracy.
\begin{enumerate}

\item As can be read from the background solution \eqref{eq:rho-rh}, all background observables such as $N_k$ are functionals of $w_\text{rh}(N)$ only through the single number $w_\text{eff}$. This is because the Friedmann-Robertson-Walker (FRW) background acts as a ``projection operator'' that maps the infinite-dimensional function space of $w_\text{rh}(N)$ onto a one-dimensional parameter space of $w_\text{eff}$. Thus, The profile of $w_\text{rh}(N)$ becomes relevant when the observable probes the operators beyond the background stress-energy tensor. More precisely speaking, the background evolution depends only on $T_{\mu\nu}$ evaluated on the homogeneous solution. However, any observables that probe local time dependence during reheating, such as horizon reentry of sub-CMB modes, perturbation evolution, or particle production depend on quantities like $H(N)$, $dH/dN$, and so on, which again depend on the local value of $w_\text{rh}(N)$, {\it not} its integral. For example, the gravitational wave spectrum $\Omega_\text{GW}h^2$ is sensitive to the profile of $w_\text{rh}$~(see e.g.~\cite{Choi:2024ilx}).

We can more formally state the above discussions using the terms of effective field theory (EFT). The background and perturbation operators can be written in the effective action as
\begin{equation}
\label{eq:Seff}
S_\text{eff} 
= 
\int d^4x \sqrt{-g} \Big[
\calL_\text{bg}(w_\text{eff}) + \calL_\text{pert}(w_\text{rh}(N), c_s(N), \cdots) + \cdots
\Big] 
\, .
\end{equation}
As the background sector depends only on $w_\text{eff}$, the functional profile of $w_\text{rh}(N)$ is invisible to the EFT truncated to homogeneous operators. We can access to the profile dependence and hence break degeneracy of $w_\text{eff}$ only when the perturbation operators are retained that resolve local time dependence during reheating.

\item Incorporating other observables, such as $\Omega_\text{GW}h^2$, along with $N_k$ is clearly one way of breaking the profile degeneracy of $w_\text{rh}$. There is yet another. We previously have assumed more than just FRW and energy conservation. We have also assumed that the post-reheating radiation bath behaves adiabatically with effectively constant relativistic degrees of freedom $g_*$. Thus, for example, once $g_*$ varies during reheating, the background reduction to $w_\text{eff}$ is no longer exact.

In writing \eqref{eq:rho-rh}, we have only assumed barotropicity $p = w_\text{rh}\rho$ during reheating, and no assumption about $g_*$ is made. But when we translate it into a relation involving temperature, we use the energy density for radiation 
\begin{equation}
\label{eq:r-energy}
\rho_r = \frac{\pi^2}{30} g_*(T) T^4 \, ,
\end{equation}
and entropy conservation
\begin{equation}
\label{eq:entropy-cons}
g_*(T) a^3T^3 = \text{constant} \, .
\end{equation}
Hence, if $g_*$ varies significantly during reheating, then
\begin{equation}
d\log{T} = - dN - \frac{1}{3} d\log g_* \, ,
\end{equation}
so that the mapping between $a$, $T$ and $\rho$ picks up explicit dependence on the local evolution of $g_*$, which cannot be absorbed into $w_\text{eff}$.

The varying $g_*$ means that additional degrees of freedom become dynamical, or that phase transition occurs, or that entropy is produced non-adiabatically and redistributed. So even at homogeneous level, the effective stress-energy tensor is no longer characterized by a single parameter $w_\text{eff}$. Thus, now the background action in \eqref{eq:Seff} depends also on $g_*(N)$, and different reheating histories with identical $w_\text{eff}$ can yield different final temperatures and different mappings between $k$ and $N_k$. In this sense, ``weak model-independence'' means setting up the boundaries up to which point the EFT description of reheating, parametrized by $w_\text{eff}$, is valid: It is so within a) a single perfect fluid, b) adiabatic evolution, and c) fixed particle contents.

\end{enumerate}

\iffalse
%
%
%
A few representative examples are as follows:
%
\begin{enumerate}

\item Varying $g_*$: Since what is really conserved is the combination $g_*a^3T^3$, the mapping between $N_k$ and $T$ is not one-to-one any longer if $g_*$ is strongly varying. In this case, we have an additional shape-sensitive term as
%
\begin{equation}
\int w_\text{rh}(N) d\log g_*(N) \, .
\end{equation}
%
We will return to this point again in an explicit simple example.

\item Delayed thermalization: If the inflaton does not decay fast enough,

\end{enumerate}
%
Our framework is not designed to extract more information from $N_k$, but to expose what information $N_k$ fundamentally cannot access. 
%
%
%
\fi

\subsection{Shape sensitivity from varying relativistic degrees of freedom}

\begin{figure}[ht]
\begin{center}
 \includegraphics[width=0.6\textwidth]{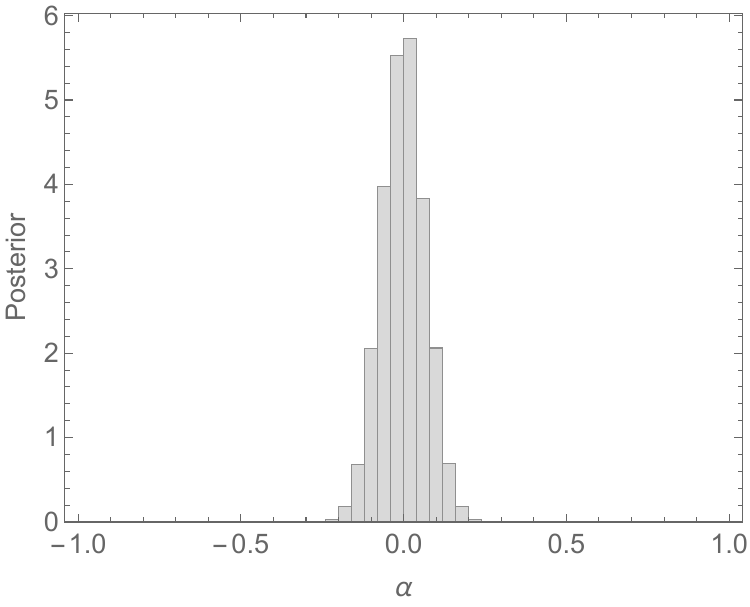}
 \caption{Posterior distribution of $\alpha$ obtained from the MCMC analysis for \eqref{eq:mcmc-toy}. We have set $N_k^\text{fid} = 55$, $A = 3.0$ and $\sigma = 0.2$. We have taken 50,000 samples with $-1 < \alpha < 1$.}
 \label{fig:mcmc}
\end{center}
\end{figure}

Before we proceed with a physically motivated simple study, we first consider a toy MCMC case for illustration how $N_k$ depends on the shape parameter $\alpha$. With a fiducial value $N_k^\text{fid}$, we add to $N_k$ the part $\delta{N}_k$ that depends on $\alpha$: 
\begin{equation}
N_k = N_k^\text{fid} + \delta{N}_k(\alpha) \, .
\end{equation}
As a minimal choice, let us consider a linear toy dependence:
\begin{equation}
\label{eq:mcmc-toy}
\delta{N}_k(\alpha) = A\alpha \, ,
\end{equation}
where $A$ is some number. We choose the likelihood of $N_k$ to be Gaussian with the variance $\sigma$, centered on the fiducial value $N_k^\text{fid}$. In Figure~\ref{fig:mcmc}, we show the posterior distribution of the shape parameter $\alpha$, chosen randomly between $-1$ and 1, obtained from the MCMC analysis with 50,000 samples incorporating \eqref{eq:mcmc-toy}. The strong Gaussian-like constraint centered near $\alpha=0$ with exponential suppression beyond $|\alpha| \sim 0.25$ clearly demonstrates that the reheating profile shape is non-degenerate and can be constrained with precise measurements, addressing the limitation of the standard parameterization using only $w_\text{eff}$.

Now, we consider an explicit toy model where $g_*$ varies as a step:
\begin{equation}
g_*(T) = \left\{
\begin{array}{ll}
g_h & (T > T_c)
\vspace{0.3em}\\
g_l & (T < T_c)
\end{array}
\right.
\, ,
\end{equation}
with $g_h > g_l$. We set up two different profiles for $w_\text{rh}(n)$ as
\begin{align}
w_a(n) & =
\left\{
\begin{array}{ll}
w_s & (0 < n < \alpha)
\vspace{0.3em}\\
w_r & (\alpha < n < 1)
\end{array}
\right.
\, ,
\\
w_b(n) & =
\left\{
\begin{array}{ll}
w_r & (0 < n < 1 - \alpha)
\vspace{0.3em}\\
w_s & (1 - \alpha < n < 1)
\end{array}
\right.
\, ,
\end{align}
with $w_s > w_r$. That is, we give different ordering of stiff and soft phases: Stiff phase comes early ($w_a$) or late ($w_b$). In both cases, the average equation of state $w_\text{eff}$ is the same:
\begin{equation}
w_\text{eff} = (w_s - w_r)\alpha + w_r \, ,
\end{equation}
so that the $\alpha$-dependence of $N_k$ can be, at least approximately, modeled by \eqref{eq:mcmc-toy}. Because of entropy conservation \eqref{eq:entropy-cons}, when $g_*$ drops from $g_h$ to $g_l$, temperature does not scale simply as $T\propto a^{-1}$, but there is an extra jump $T \propto g_*^{-1/3}a^{-1}$. This makes the background mapping ordering-sensitive, not just average-sensitive.

First we begin with the energy density evolution \eqref{eq:rho-rh}. Since $w_\text{eff} = \calW$ is identical for both profiles, the energy density evolves identically so we have the same $\rho_\text{rh}$. Up to here, it is still degenerate. Now suppose that the temperature crosses $T_c$ at some $e$-folds $N_c$ during reheating. From \eqref{eq:r-energy}, at the crossing we have
\begin{equation}
\rho(N_c) = \frac{\pi^2}{30} g_h T_c^4 \, .
\end{equation}
Because $\rho$ depends on the integrated history up to that time, the value of $N_c$ depends on how much stiff versus soft expansion happened before the crossing as during the stiff phase $\rho$ decreases faster. So, If the stiff phase happens early ($w_a$), the energy density drops faster before hitting $T_c$, and $N_c$ occurs earlier. If, conversely, the stiff phase comes late ($w_b$), the crossing happens later. Thus, even though $w_\text{eff}$ is degenerate, the moment of crossing $N_c$ becomes different depending on the profile of $w_\text{rh}$. This gives rise to different pair $(a_\text{rh},T_\text{rh})$ for different profiles.

\section{Conclusions}
\label{sec:conc}

In this article, we have studied the total expansion a certain mode experiences during inflation since the moment of horizon crossing until the end of inflation. In doing so, we have shown that the contributions which are dependent on the detailed dynamics of inflation and reheating are separated from those which are not. Furthermore, by assuming that the equation of state during the reheating stage $w_\text{rh}$ is a function of $e$-folds, its effects, i.e. the speculative dynamics during reheating, are confined within a time integral, giving rise to a simple function of model-dependent parameters with magnitude $\calO(1)$.

We have illustrated the total expansion of certain $k$-modes by two choices of the equation of state during reheating, \eqref{eq:w-rh1} and \eqref{eq:w-rh3}. Depending on the parameter $\alpha$ that controls the shape of $w_\text{rh}$ and the reheating temperature $T_\text{rh}$, the number of $e$-folds can be specified experienced by the $k$-mode. We could also confirm the two well-known observations that 1) slower reheating gives a smaller value of $N_k$ and, conversely, 2) a larger value of $N_k$ is obtained for $w_\text{rh} \geq 1/3$.

Moreover, we have clarified on very general ground when the degeneracy in the shape profile of $w_\text{rh}$ matters. There are two ways of breaking degeneracy. One is to incorporate another perturbation observable, and the other is to vary the relativistic degrees of freedom $g_*$. We have shown how degeneracy can be broken by presenting an explicit example of varying $g_*$.

Therefore, by an appropriate modeling of the equation of state during reheating as a function of $e$-folds and by choosing the desired reheating temperature, the impacts of the unknown dynamics of reheating on the expansion of perturbation modes can be easily estimated. For any generic models of reheating it is possible to parametrize $w_\text{rh}$ as a function of $e$-folds even if we cannot derive the analytic form of $w_\text{rh}$ microscopically, e.g. by numerical fitting. Thus, we believe our approach can be universally adopted for generic studies of various aspects of reheating epoch.

\subsection*{Acknowledgements}

We thank Kiwoon Choi, Suro Kim, Ryo Saito, Tomo Takahashi, Yuki Watanabe and Chul-Moon Yoo for helpful comments.
We are also grateful to the anonymous editor for her/his constructive feedback that improved the contents of this article.
This work is supported in part by Basic Science Research Program through the National Research Foundation of Korea (RS-2024-00336507). JG also acknowledges the Ewha Womans University Research Grant of 2025 (1-2025-0739-001-1). 
JG is grateful to the Asia Pacific Center for Theoretical Physics for hospitality during this work was under progress.

\end{document}